\documentclass[preprint,12pt]{elsarticle}

\usepackage{graphicx}

\usepackage{amsmath,amssymb,amsfonts,amsthm,enumerate}

\usepackage{ifthen}
\usepackage{hyperref}
\usepackage{subfigure}

\theoremstyle{plain}

\newtheorem{proposition}{Proposition}
\newtheorem{corollary}{Corollary}

\theoremstyle{definition}

\theoremstyle{remark}
\newtheorem{remark}{Remark}
\newtheorem{example}{Example}

\newcommand{\R}{\mathbb{R}}
\newcommand{\cS}{{\cal S}}

\newcommand{\cK}{{\cal K}}
\newcommand{\cA}{{\cal A}}
\newcommand{\GK}{\Gamma_K}
\newcommand{\GKs}{\Gamma^*_K}
\newcommand{\bGKs}{\bar{\Gamma}^*_K}
\newcommand{\indicator}[1]{\mathbf{1}_{\left\{ {#1} \right\} }}

\newcommand{\cN}{{\cal N}}
\newcommand{\bq}{\mathbf{q}}
\newcommand{\bw}{\mathbf{w}}
\newcommand{\Integers}[0]{\mathbb{Z}}
\newcommand{\defined}{\stackrel{\triangle}{=}}

\journal{Theoretical Computer Science}

\begin{document}

\begin{frontmatter}


\title{Sensor Networks: from Dependence Analysis Via Matroid Bases to Online Synthesis\tnoteref{label1,supp}}
\tnotetext[label1]{An early version of the work appeared in the $7$th International Symposium on 
Algorithms for Sensor Systems, Wireless Ad Hoc Networks and Autonomous Mobile Entities, ALGOSENSORS 2011, Saarbruecken, Germany.}
\tnotetext[supp]{This work was partially supported by MAFAT, the Chief Scientist (Magnet - Captain) and ISF.}

\author[cse]{Asaf Cohen\corref{cor1}}
\cortext[cor1]{Corresponding author. Phone: 972-8-6428062.}
\ead{coasaf@bgu.ac.il}
\ead[url]{www.bgu.ac.il/~coasaf}

\author[cs]{Shlomi Dolev}
\ead{dolev@cs.bgu.ac.il}
\ead[url]{www.cs.bgu.ac.il/~dolev}

\author[cs,ashkelon]{Guy Leshem}
\ead{leshemg@cs.bgu.ac.il}

\address[cse]{Dept. of Communication Systems Eng., Ben-Gurion University,
Beer-Sheva, Israel.}
\address[cs]{Dept. of Computer Science, Ben-Gurion University,
Beer-Sheva, Israel.}
\address[ashkelon]{Dept. of Computer Science, Ashkelon Academic College,
Ashkelon, Israel.}
\begin{abstract}
Consider the two related problems of sensor selection and sensor fusion. In the first, given a set of sensors, one wishes to identify a subset of the sensors, which while small in size, captures the essence of the data gathered by the sensors. In the second, one wishes to construct a fused sensor, which utilizes the data from the sensors (possibly after discarding dependent ones) in order to create a single sensor which is more reliable than each of the individual ones.

In this work, we rigorously define the dependence among sensors in terms of joint empirical measures and incremental parsing. We show that these measures adhere to a polymatroid structure, which in turn facilitates the application of efficient algorithms for sensor selection. We suggest both a random and a greedy algorithm for sensor selection. Given an independent set, we then turn to the fusion problem, and suggest a novel variant of the exponential weighting algorithm. In the suggested algorithm, one competes against an augmented set of sensors, which allows it to converge to the best fused sensor in a family of sensors, without having any prior data on the sensors' performance.   
\end{abstract}


\end{frontmatter}

\section{Introduction}\label{sec. intro}
Sensor networks are used to gather and analyze data in a variety of applications. In this model, numerous sensors are either spread in a wide area, or simply measure different aspects of a certain phenomenon. The goal of a central processor which gathers the data is, in general, to infer about the environment the sensors measure and make various decisions. 
An example to be kept in mind can be a set of sensors monitoring various networking aspects in an organization (incoming and outgoing traffic, addresses, remote procedure calls, http requests to servers and such). In many such cases, an anomalous behavior detected by a single sensor may not be reliable enough to announce the system is under attack. Moreover, different sensors might have correlated data, as they measure related phenomenons. Hence, the central processor faces two problems. First, how to identify the set of sensors which sense independent data, and discard the rest, which only clutter the decision process. The second, how to intelligently combine the data from the sensors it selected in order to decide whether to raise an alarm or not. 

In this work, we target both problems. First, we consider the problem of sensor selection. Clearly, as data aggregated by different sensors may be highly dependent, due to, for example, co-location or other similarities in the environment, it is desirable to identify the largest set of independent (or nearly independent) sensors. This way, sensor fusion algorithms can be much more efficient. For example, in the fusion algorithm we present, identifying the set of independent sensors allows us to create families of fused sensors based on fewer sensors, hence having a significantly smaller parameter space.
 Moreover, identifying independent sensors is of benefit also to various control methods, were a few representative independent inputs facilitate easier analysis. Note that the sensor selection problem is different from the data compression problem, where the dependence among the data sets is reduced via some kind of Slepian-Wolf coding \cite{SlepianWolf73}. Herein, we do not wish all data to be reconstructed at the center, but focus only identify good sets of independent sensors, such that \emph{their} data can be analyzed, disregarding other sensors. In other words, we do not wish to replace Slepian-Wolf coding by sending data of independent sensors, only identify the independent subsets. For example, the randomized algorithm we suggest gathers data only from small subsets of the sensors, yet is assured to identify independent sets with high probability. In a similar manner, A greed algorithm we suggest can identify subset of sensors with relatively high independence among them (compared to other subsets), even in cases we do not wish to identify a subset containing \emph{all} the information.  

Given two data sets, a favored method to measure their dependence is through various \emph{mutual information estimates}. Such estimates arise from calculating marginal and joint empirical entropies, or the more efficient method of incremental (Lempel-Ziv) parsing \cite{Ziv_Lemp78}. Indeed, LZ parsing was used, for example, for multidimensional data analysis \cite{zozor2005lempel}, neural computation \cite{blanc2008quantifying} and numerous other applications. However, although the ability of the parsing rule to approximate the true entropy of the source, and hence, as one possible consequence, identify dependencies in the data, applications reported in the current literature were ad-hoc, using the resulting measures to compare between mainly \emph{pairs of sources}. 

To date, there is no rigorous method to analyze independence among large sets, and handle cases where one sensor's data may depend on measurements from many others, including \emph{various delays}. In this work, we give the mathematical framework which enables us to both rigorously define the problem of identifying sets of independent sources in a large set of sensors and give highly efficient approximations algorithms based on the observations we gain.
 
Still, when no single sensor is reliable enough to give an accurate estimate of the phenomenon it measures, sensor fusion is used \cite{hall1997introduction,sasiadek2002sensor,waltzdata}. In the second part of this work, we consider the problem of sensor fusion. In this case, for a given set of sensors, one wishes to generate a \emph{new sensor}, whose performance over time is (under some measure) better than any single sensor in the set. Clearly, in most cases, choosing the best-performing sensor in the set might not be enough. We wish, in general, to create a sensor whose performance is strictly better than any given sensor in the original set, by utilizing data from several sensors simultaneously and intelligently combining it. 
\subsection{Contribution.}
Our main contributions are the following. First, we show how to harness the wide variety of algorithms for identifying largest independent sets in matroids, or the very related problems of minimum cycle bases and spanning trees in graphs to our problem of identifying sets of independent sensors in a sensor network. Our approach is based on highly efficient (linear time in the size of the data) methods to estimate the dependence among the sensors, such as the Lempel-Ziv parsing rule \cite{Ziv_Lemp78}. The key step, however, is in showing how these estimates can either yield a polymatroid, or at least approximate a one, thus facilitating the use of polynomial time algorithms to identify the independent sets, such as \cite{karger1993random,berger2004minimum}. We construct both random and a greedy selection algorithms, and analyze their performance. 

We then turn to the problem of (non-correlated) sensor fusion. In particular, we describe an online fusion algorithm based on exponential weighting \cite{Vovk90}.
While weighted majority algorithms were used in the context of sensor networks \cite{jeon1999decision,yu2006learning,polikar2006multiple}, in these works, the exponential weighting was used only to identify good sensors and order them by performance. Hence, applied directly, this algorithm does not yield a good \emph{fused} sensor. In this part of our work, we suggest a novel extension by creating parametric families of synthesised sensors. This way, we are able to \emph{span a huge set of fused sensors}, and choose online the best fused sensor. That is, given a set of sensors $\cS$, this algorithm \emph{constructs synthesised sensors}, from which it selects a sensor whose performance converges to that of the best sensor in both $\cS$ and the constructed parametric family of synthesised sensors. In other words, this algorithm results in online sensor fusion.

We rigorously quantify the regret of the suggested algorithm compared to the best fused sensor. In this way, a designer of a sensor fusion algorithm has a well-quantified trade-off: choosing a large number of parameters, thus covering more families of fusion possibilities, at the price of higher regret.   
\section{Preliminaries}\label{sec. prelim}
The basic setting we consider is the following. A set of sensors, $\cS = \{S_1,\ldots,S_{K}\}$ is measuring a set of values in a certain environment. Each sensor may depend on a different set of values, and may base its decision on these values in a different way. However, each sensor, at each time instance, estimates whether a target exists in the environment or not. Thus, the input to sensor $S_j$ at time $t$ is some vector of measurements $V^j_t$, based on which it will output a value $p^j_t \in [0,1]$, which is his estimate for the probability a target exists at time $t$.\footnote{The model of target identification and the requirement for $p^j_t \in [0,1]$ is mainly for illustration purposes. The algorithms we describe are suitable for different sets as well, with the adequate quantization or scaling.} Throughout, capital letters denote random variables while lower case denotes realizations. Hence, $P^j_t$ denotes the possibly random output of sensor $S_j$, $j=1,2,\ldots,K$ at time $t$, $t=1,2,\ldots,n$.  

Let $\{x_t\}_{t=1}^{n}$ be the binary sequence indicating whether a target actually appeared at time $t$ or not. The normalized cumulative loss of the sensor $S_j$ over $n$ time instances is defined as
\begin{equation}\label{eq. comu. loss}
L_{S_j}(x_1^n)=\frac{1}{n}\sum_{t=1}^{n}d(p^j_t,x_t),
\end{equation}
for some distance function $d:[0,1]\times\{0,1\} \mapsto \R$. If the sensor's output is binary (a sensor either decides a target exists or not), then $p^j_t \in \{0,1\}$ and a reasonable distance measure is the Hamming distance, that is, 
\[
d(a,b) = \left\{ \begin{array}{ll}
0 & \textrm{if $a=b$}\\ 1 & \textrm{if $a \ne b$}.
\end{array} \right.
\]
If the sensor's output is in $[0,1]$, then we think of it as the sensor's estimate for the probability a target exists, and a reasonable $d$ is the \emph{log-loss}, 
\[
d(p,x) = -x\log(p)-(1-x)\log(1-p).
\]
In any case, the goal of a good sensor $S$ is to minimize the normalized cumulative loss $L_S(x_1^n)$, as given by \eqref{eq. comu. loss}. Roughly speaking, in the first part of this paper, we wish to use the estimates $\{P^j_t\}$ to identify dependencies among the sensors, and in the second part we wish to construct fused sensors, whose cumulative loss is lower than any given sensor.
\subsection{Polymatroids, Matroids and Entropic Vectors.}
Let $\cK$ be an index set of size $K$ and $\cN$ be the power set of $\cK$. A function $g:\cN \mapsto \R$ defines a \emph{polymatroid} $(\cK,g)$ with a ground set $\cK$ and rank function $g$ if it satisfies the following conditions \cite{Oxley92}:
\begin{eqnarray}
g(\emptyset)&=&0,\label{eq. poly axioms 1}
\\
g(I) &\leq& g(J) \text{ for } I \subseteq J \subseteq \cK,\label{eq. poly axioms 2}
\\
g(I) + g(J) &\ge& g(I \cup J) + g(I \cap J) \text{ for } I,J \subseteq \cK.\label{eq. poly axioms 3}
\end{eqnarray}
For a polymatroid $g$ with ground set $\cK$, we represent $g$ by the vector $(g(I):I\subseteq \cK) \in \R^{2^K-1}$ defined on the ordered, non-empty subsets of $\cK$. We denote  
the set of all polymatroids with a ground set of size $K$ by $\GK$. Thus $\bw \in \GK$ if and only if $w(I)$ and $w(J)$ satisfy equations \eqref{eq. poly axioms 1}--\eqref{eq. poly axioms 3} for all $I,J \subseteq \cK$, where $w(I)$ is the value of $\bw$ at the entry corresponding to the subset $I$. If, in addition to \eqref{eq. poly axioms 1}--\eqref{eq. poly axioms 3}, $g(\cdot)\in \Integers^+$ and $g(I) \leq |I|$, then $(\cK,g)$ is called a \emph{matroid}.

Now, assume $\cK$ is some set of discrete random variables. For any $A \subseteq \cK$, let $H(A)$ denote the joint entropy function. An entropy vector $\bw$ is a $(2^K-1)$-dimensional vector whose entries are the joint entropies of all non-empty subsets of $\cK$. It is well-known that the entropy function is a polymatroid over this ground set $\cK$. Indeed, \eqref{eq. poly axioms 1}--\eqref{eq. poly axioms 3} are equivalent to the Shannon information inequalities \cite{Yeung02}. However, there exists points $\bw \in \GK$ ($K>3$) for which there is no set of $K$ discrete random variables whose joint entropies equal $\bw$. Following \cite{Chan_Grant08}, we denote by $\GKs$ the set of all $\bw \in \GK$ for which there exists at least one random vector whose joint entropies equal $\bw$. A $\bw\in\GKs$ is called \emph{entropic}. 
Finally, denote by $\bGKs$ the convex closure of $\GKs$. Then $\bGKs=\GK$ for $K \leq 3$ but $\bGKs \ne \GK$ for $K > 3$ \cite{Yeung02}. 
\section{A Matroid-Based Framework for Identifying Independent Sensors}\label{sec. matroid based}
In this section, we use the incremental parsing rule of Lempel and Ziv \cite{Ziv_Lemp78} to estimate the \emph{joint empirical entropies} of the sensors' data. We then show that when the sensors data is stationary and ergodic, the vector of joint empirical entropies can be approximated by some point in the polyhedral cone $\GK$. In fact, this point is actually in $\bGKs$. As asymptotically entropic polymatroids are well approximated by asymptotically entropic matroids \cite[Theorem 5]{Matus07}, the point in $\R^n$ which corresponds to the joint empirical entropies of the sensors is approximated by the \emph{ranks of some matroid}. This enables us to identify independent sets of sensors, and, in particular, largest independent sets, by identifying the bases (or circuits) of the matroid.  
Doing this, the most complex dependence structures among sensors, including both dependence between past/future data and dependence among values at the same time instant can be identified. Non-linear dependencies are also captured.

We now show how to approximate an entropy vector (hence, a polymatroid) for the sensor data. We prove that indeed for large enough data and ergodic sources the approximation error is arbitrarily small. This polymatroid will be the input from which we will identify the independent sensors.

We first consider the most simple case in which one treats the sensors as having memoryless data. That is, sensors for which each reading (in time) is independent of the previous or future readings. Note, however, that this model still allows the reading of a sensor to depend on the readings of other sensors \emph{at that time instant}. The dependence might be a simple (maybe linear) dependence between two sensors, or a more complex one, where one sensor's output is a random function of the outputs of a few others. It is important to note that it is inconsequential if the sensors are indeed memoryless or not. Using this simplified method, only dependencies across a single time instant will be identified. A generalization for time-dependent data appears in the next sub-section.

For the sake of simplicity, assume now all $P^j_t$ are binary. Given a sequence $\{p_i\}_{i=1}^n$, denote by $N(0|\{p_i\}_{i=1}^n)$ and $N(1|\{p_i\}_{i=1}^n)$ the number of zeros and ones in $\{p_i\}_{i=1}^n$, respectively. That is, 
\[
N(0|\{p_i\}_{i=1}^n) \defined \sum_{i=1}^n \indicator{p_i=0},
\]
where $\indicator{\cdot}$ is the indicator function. When the sequence indices are clear from the context, we will abbreviate this by $N(0|p)$. Hence, 
\[
T^n_p \defined \left(\frac{1}{n}N(0|p),\frac{1}{n}N(1|p)\right)
\]
denotes the \emph{type} of the sequence $p$, that is, its empirical frequencies \cite{cover2006elements}.

In a similar manner, we define the empirical frequencies of several sequences together, e.g. pairs. For example, 
\[
N\left(0,1|p^1,p^2\right) \defined \sum_{i=1}^n \indicator{p^1_i=0, p^2_i=1}.
\]
In this case, the $4$-tuple 
\[
T^n_{p^1,p^2} \defined \left(\frac{1}{n}N\left(0,0|p^1,p^2\right),\frac{1}{n}N\left(0,1|p^1,p^2\right),\frac{1}{n}N\left(1,0|p^1,p^2\right),\frac{1}{n}N\left(1,1|p^1,p^2\right)\right)
\]
denotes the \emph{joint type} of $(p^1,p^2)$, hence, it includes the empirical frequencies of the two sequences \emph{together}, over their product alphabet $\{0,1\} \times \{0,1\}$. For more than two sequences, we denote by $T^n_{p^1,\ldots,p^\cS}$ the joint type of the sequences $(p^j, j \in \cS)$.

For a probability vector $\bq = (q_1,\ldots,q_m)$, let $H(\bq)$ denote its entropy, that is,
\[
H(\bq) = -\sum_{i=1}^{m}q_i\log(q_i).
\]
Let $\bw_n$ be the $(2^K-1)$-dimensional vector whose entries are all the joint \emph{empirical entropies} calculated from $\{(p^j_t)_{j\in \cS}\}_{t=1}^{n}$. I.e,
\[
\bw_n \defined \big( H(T^n_{p^1}),\ldots, H(T^n_{p^K}),
H(T^n_{p^1,p^2}),H(T^n_{p^1,p^3}), \ldots , H(T^n_{p^1,\ldots,p^K}) \big)
\]
Under these definitions, we have the following.
\begin{proposition}\label{prop. wn in gns}
For every realization of the sensors' data, $\bw_n \in \GKs$.
\end{proposition}
\def\PROOF1{
\begin{proof}
We wish to show that the vector of joint empirical entropies, $\bw_n$, is entropic for any finite $n$. Hence, $\bw_n \in \GKs$. The important observation is that empirical measures (as defined herein) are legitimate probability measures (even if the approximation error compared to the true measure is large), hence entropies calculated from them give rise to an entropic polymatroid.

Since for any subset $I$, $T^n_{p^j, j\in I}$ clearly defines a valid distribution (all entries are in $[0,1]$ and they sum up to $1$), consistency is the only property it remains to show. Assume all sensors' output belong to some finite alphabet $\cA$. We have
\begin{eqnarray*}
\sum_{a_1 \in \cA} \frac{1}{n} N(a_1,\ldots,a_K|p^1,\ldots,p^K) &=& \frac{1}{n}\sum_{a_1 \in \cA}\sum_{i=1}^n \indicator{p^1=a_1,\ldots,p^K=a_K}
\\
&=& \frac{1}{n}\sum_{i=1}^n \indicator{p^2=a_2,\ldots,p^K=a_K}
\\ &=& \frac{1}{n}N(a_2,\ldots,a_K|p^2,\ldots,p^K)
\end{eqnarray*}
which completes the proof.
\end{proof}
} 
\PROOF1
Let $\bw$ denote the true (memoryless) entropy vector of the sources. That is,
\[
\bw \defined \big( H(P_1^1),\ldots ,H(P_1^K),
H(P_1^1,P_1^2),H(P_1^1,P_1^3), \ldots, H(P_1^1,\ldots,P_1^K) \big).
\]
For stationary and ergodic sources, the following Proposition is a direct application of Birkhoff's ergodic theorem. 
\begin{proposition}\label{prop. empirical antropy to true one}
Let $\{(p^j_t)_{j\in \cS}\}_{t=1}^{n}$ be drawn from a stationary and ergodic source $\{(P^j_t)_{j\in \cS}\}_{t=1}^{\infty}$ with some probability measure $Q$. Then, for any subset $\cS' \subseteq \cS$, we have $\lim_{n\to\infty}H(T^n_{p^j, j \in \cS'}) = H\left((P^j_1)_{j\in \cS'}\right)$  $Q$-a.s. (almost surely). As a result,
$\Pr\left(\lim_{n \to\infty} \bw_n = \bw \right)=1$.
\end{proposition}
That is, the entropy calculated from the empirical distribution converges to the true entropy. Moreover, the vector of empirical entropies converges almost-surely (a.s.) to the true entropy vector, which is, of course, an entropic polymatroid. To be able to harness the diverse algorithmic literature on matroids (such as matroid optimization relevant for our independence analysis application), we mention that by \cite[Theorem 5]{Matus07}, describing the cone of asymptotically entropic polymatroids, $\bGKs$, is reduced to the problem of describing asymptotically entropic \emph{matroids}.  
\subsection{Dependence Measures for Sensors with Memory.}
Till now, we considered sensors for which the data for any \emph{individual} sensor is a stationary and ergodic process, yet, through first-order empirical entropies, only the dependence along a single time instant was estimated. While being very easy to implement (linear in the size of the data), this method fails to capture complex dependence structures. For example, consider a sensor whose current data depends heavily on \emph{previous data} acquired by \emph{one or several other} sensors. 

To capture dependence in time, we offer the incremental parsing rule \cite{Ziv_Lemp78} as a basis for an empirical measure. We show that indeed such a measure will converge almost surely to a polymatroid, from which maximal independent sets can be approximated. We start with a few definitions.

Let $\{p_i\}_{i=1}^{n}$ be some sequence over a finite alphabet of size $\alpha$. The ZL78 \cite{Ziv_Lemp78} parsing rule is a sequential procedure which parses the sequence $p$ in a way where a new phrase is created as soon as the still unparsed part of the string differs from all preceding phrases. For example, the string 
\[
0100011011000001010011\ldots
\]
is parsed as 
\[
0,1,00,01,10,11,000,001,010,011,\ldots.
\]
Let $c(\{p_i\}_{i=1}^n)$ denote the number of distinct phrases whose concatenation generates $\{p_i\}_{i=1}^n$. Furthermore, let $\rho_E(s)(\{p_i\}_{i=1}^n)$ denote the compression ratio achieved by the best finite-state encoder with at most $s$ state, and define
\[
\rho(p) = \lim_{s\to\infty}\limsup_{n\to\infty}\rho_{E(s)}(\{p_i\}_{i=1}^n).
\]
In a nutshell, the main results of \cite{Ziv_Lemp78} states that
on the one hand
\[
\rho(p) \ge \limsup_{n\to\infty}\frac{c(\{p_i\}_{i=1}^n)\log c(\{p_i\}_{i=1}^n)}{n \log \alpha}
\]
where $\alpha$ is the alphabet size. On the other hand, for any sequence $\{p_i\}_{i=1}^n$, there exists a finite state encoder with a compression ratio $\rho_E(\{p_i\}_{i=1}^n)$ satisfying
\[
\rho_E(\{p_i\}_{i=1}^n) \leq \frac{c(\{p_i\}_{i=1}^n)+1}{n \log \alpha}\log\left(2\alpha(c(\{p_i\}_{i=1}^n)+1)\right).
\]
Thus,
\[
H^{LZ}(\{p^1_i\}_{i=1}^{n}) \defined \frac{c(\{p^1_i\}_{i=1}^{n})\log c(\{p^1_i\}_{i=1}^{n})}{n\alpha}
\]
is an asymptotically attainable lower bound on the compression ratio $\rho(p)$. Denote by $\bar{H}(P)$ the \emph{entropy rate} of a stationary source $P$, that is,\\ $\lim_{n\to\infty}\frac{1}{n}H(P_1,\ldots,P_n)$. For $K$ sources $P^1,\ldots,P^K$, the entropy rate vector $\bar{\bw}$ is defined as
\[
\bar{\bw} \defined \big( \bar{H}(P^1),\ldots, \bar{H}(P^K),
\bar{H}(P^1,P^2),\bar{H}(P^1,P^3), \ldots, \bar{H}(P^1,\ldots,P^K) \big).
\]
Analogously to the memoryless case, herein we also define the joint parsing rule in the trivial way, that is, parsing any subset of $1<k\leq K$ sequences as a single sequence over the product alphabet. Define the LZ-based estimated entropy vector $\bw^{LZ}_n$ as (suppressing the dependence on $n$)
\[
\bw^{LZ}_n \defined \big( H^{LZ}(p^1),\ldots, H^{LZ}(p^K),
H^{LZ}(p^1,p^2), \ldots, H^{LZ}(p^1,\ldots,p^K) \big).
\]
The following is the analogue of Proposition \ref{prop. empirical antropy to true one} for the non-memoryless case.
\begin{proposition}\label{prop. convergence of wLZ}
Let $\{(p^j_t)_{j\in \cS}\}_{t=1}^{n}$ be drawn from a stationary and ergodic source $\{(P^j_t)_{j\in \cS}\}_{t=1}^{\infty}$. Then, $\bar{\bw} \in \bGKs$ and we have 
\[
\Pr\left(\lim_{n \to\infty} \bw_n^{LZ} =\bar{\bw} \right)=1.
\]
\end{proposition} 
\def\PROOF2{
\begin{proof}
We wish to see that $\bw_n^{LZ}$ converges to $\bar{\bw}$, and that indeed $\bar{\bw} \in \bGKs$. It is not hard to show that $\bar{\bw} \in \bGKs$. To see this, remember that $H(\{P^j_i\}_{i=1}^{n}, j \in I)$, ranging over all subsets $I\subseteq \cS$ forms an entropic polymatroid \cite{Yeung02}. Hence $\frac{1}{n}H(\{P^j_i\}_{i=1}^{n}, j \in I)$ forms an asymptotically entropic polymatroid (as the closure of the entropic region is convex), hence $\bar{\bw} \in \bGKs$.

 In \cite{Ziv_Lemp78}, it was proved that for stationary and ergodic sources each entry in $\bw_n^{LZ}$ converges to the true entropy rate. That is,
\[
\lim_{n \to \infty} H^{LZ}(P^j, j \in I) = \lim_{n \to \infty}\frac{1}{n}H(\{P^j_i\}_{i=1}^{n}, j \in I) \quad \text{a.s.}
\]
Since we have only finitely many entries in $\bw_n^{LZ}$, a simple union bound gives Proposition \ref{prop. convergence of wLZ}. 
\end{proof}
} 
\PROOF2

Note, however, that the analogue of Proposition \ref{prop. wn in gns} is not true in this case. That is, for finite $n$, $\bw_n^{LZ}$ might not satisfy the polymatroid axioms at all. Nevertheless, by Proposition \ref{prop. convergence of wLZ}, for large enough $n$, $\bw_n^{LZ}$ is sufficiently close to $\bGKs$. A fortiori, it is sufficiently close to $\GK$.  
Moreover, for ergodic sources with finite memory, namely, sources for which
\begin{multline*}
\Pr(P_n=a_n|P_{n-1}=a_{n-1}, P_{n-2}=a_{n-2}, \ldots) \\ =\Pr(P_n=a_n|P_{n-1}=a_{n-1}, \ldots, P_{n-m}=a_{n-m})
\end{multline*}
for some finite $m$, there exist a few strong tail bounds on the probability that the LZ compression ratio exceeds a certain threshold. For example, if $\|\bw\|_0$ denotes the maximal entry in $\bw$, we have the following proposition.
\begin{proposition}\label{prop. for markov sources}
Let $\{(p^j_t)_{j\in \cS}\}_{t=1}^{n}$ be drawn from a stationary and ergodic Markov source $\{(P^j_t)_{j\in \cS}\}_{t=1}^{\infty}$. Then, with probability at least $1-O(\frac{2^K-1}{\sqrt{n}})$, $\|\bw_n^{LZ} - \bar{\bw}\|_0 \leq \frac{\bar{H}\left(P^j, j\in \cS \right)}{\log n}$.
\end{proposition}
\def\PROOF3{
\begin{proof}
By \cite[Corollary 2]{savari1997redundancy}, we have
\[
\Pr\left(|\bw_n^{LZ}(1)-\bar{H}(P^1)| > \frac{\bar{H}(P^1)}{\log n}\right) = O\left(\frac{1}{\sqrt{n}}\right).
\]
Remembering that $\bar{H}(P^i, i\in I) \leq \bar{H}(P^i, i\in \cS)$ for any $I \subseteq \cS$ and using the union bound on all entries of $\bw_n^{LZ}$ results in 
\begin{multline*}
\Pr\left(\bigcup_{I \subseteq \cS}\left\{|\bw_n^{LZ}(I)-\bar{H}(P^i, i\in I)| > \frac{\bar{H}\left(P^j,j\in \cS\right)}{\log n}\right\}\right)
\\
\leq 
\Pr\left(\bigcup_{I \subseteq \cS}\left\{|\bw_n^{LZ}(I)-\bar{H}(P^i, i\in I)| > \frac{\bar{H}\left(P^j,j\in I\right)}{\log n}\right\}\right)
\\ = O\left(\frac{2^K-1}{\sqrt{n}}\right)
\end{multline*}
which completes the proof.
\end{proof}
} 
\PROOF3
The usefulness of Proposition \ref{prop. for markov sources} is twofold. First, it gives a practical bound on the approximation the vector $\bw_n^{LZ}$ gives to $\bar{\bw}$. However, assume $\bar{\bw}$ is a matroid. This is the case, for example, when bits in the sensors' data are either independent or completely dependent (in fact, in this case $\bar{\bw}$ is a \emph{linearly representable} binary matroid). Since $\bw_n^{LZ}$ might not satisfy the polymatroid axioms at all, using Proposition \ref{prop. for markov sources} one can then easily check when can the entries of $\bw_n^{LZ}$ be rounded to the nearest integer in order to achieve $\bar{\bw}$ exactly.
\begin{remark} We mention that a different approach to target sensors with memory is to calculate \emph{high order} empirical entropies, that is, entropies calculated from the frequency count of the data seen by a sliding window of a fixed length $l$. With this approach, the achieved vector is entropic (hence a polymatroid) for any finite $n$. Moreover, with a good tail bound such as \cite{lezaud1998chernoff} for irreducible Markov chains over a finite alphabet, we are able to show fast convergence to the true vales. The complexity, however, grows exponentially with $l$. Thus, approaching entropy rates in order to capture long-time dependencies is of exponential complexity. In the LZ method we suggest, while the alphabet size indeed grows exponentially, complexity is a function of $\log$ the alphabet size.
\end{remark}
\section{Identifying Independent Sets of Sensors.}
When the number of sensors is small, and the complexity of calculating all entries of $\bw_n^{LZ}$ is reasonable, one can find a subset with high enough entropy (strong independence) by simply taking the smallest set of sensors with high enough $\frac{c \log c}{n \alpha}$. However, when the number of sensors is larger (even a few dozens), this method is prohibitively complex, and more suffisticated algorithms (and their analysis) are required. 

Thus, having set the ground, in this section we utilize optimization algorithms for submodular functions, and matroids in particular, in order to find maximal independent sets of sensors efficiently. Herein, we include two examples: a random selection algorithm, which fits cases where true data forms a matroid, for which possibly many subsets of sensors include the desired data, and a greedy algorithm, which easily fits any dependence structure (while matroids asymptotically span the entropic cone, an additional approximation step is required \cite{Matus07}). It is important to note that, unlike the greedy selection (also used in \cite{shamaiah2010greedy} in the context of maximum a posteriori estimates) which approximates the optimum value \emph{up to a constant factor}, the random selection process we suggest here can guarantee exact approximation.
\begin{figure*}[ht]
\begin{center}
\framebox{\parbox{\textwidth}{
{\bf Algorithm  \sf RandomSelection \\}
\% Input: A set of $\cS$ sensors. A parameter $0\leq q \leq 1$. \\
\% Output: A subset $I\subseteq \cS$, of expected size $qK$, which with high probability contains a maximal independent set of $\cS$ (see conditions in Corollary \ref{cor. random selection}).
\begin{itemize}
\item Include a sensor $j$ in subset $I$ with probability $q$, independently of the other sensors.
\end{itemize}
}}
\end{center}
\end{figure*}
 
The randomized algorithm is given in Algorithm {\sf RandomSelection}. As simple as it looks, by Proposition \ref{prop. for markov sources} and \cite[Theorem 5.2]{karger1993random}, under mild assumptions on the true distribution of the data, it guarantees that indeed with high probability such a random selection produces a subset of sensors which is a $q$-fraction of the original, yet if the original contains enough bases (maximal independent sets), then the subset contains a base as well. This is summarized in the following corollary.
\begin{corollary}\label{cor. random selection}
Let $\{(p^j_t)_{j\in \cS}\}_{t=1}^{n}$ be drawn from a stationary and ergodic Markov source. Assume that $\bar{\bw}$ is a matroid of rank $r$ which contains $a+2+\frac{1}{q}\ln r$ disjoint bases. Then, with probability at least $1-e^{-aq}-O(\frac{2^K-1}{\sqrt{n}})$, the subset $I$ produced by Algorithm {\sf RandomSelection} contains a maximal independent set of sensors.
\end{corollary}
\begin{figure*}[ht]
\begin{center}
\framebox{\parbox{\textwidth}{
{\bf Algorithm  \sf GreedySelection \\}
\% Input: Data of $K$ sensors, $\{p^1_t,\ldots,p^K_t\}_{t=1}^{n}$. \\
\% Output: At each time instant, a set $I$ of sensors.
\begin{itemize}
\item Initialization: $I=\phi, \hat{H}=0$.
\end{itemize}
\begin{enumerate}
\item\label{maximize} $j^* = \text{argmax}_{j\notin I} \bw_n^{LZ}(I \cup \{j\})$
\item if $\bw_n^{LZ}(I \cup \{j^*\}) > \hat{H}$ 
\begin{itemize}
\item $I \gets I \cup \{j^*\}$, then $\hat{H} \gets \bw_n^{LZ}(I)$
\item Go to step \ref{maximize}.
\end{itemize}
\end{enumerate}
}}
\end{center}
\end{figure*}
At first sight, Algorithm {\sf RandomSelection} does not depend on any of the discussed dependence measures in this paper. Yet, it power \emph{is drawn from them}: once we have established the estimated entropy vector as the key variable in determining dependence, we know that this asymptotic matroid is the one we should analyze for independent sets, \emph{according to its features we should choose the parameters} in {\sf RandomSelection} and these features will indeed eventually determine the success probability of {\sf RandomSelection}.  

On the other hand, algorithm {\sf GreedySelection} takes a different course of action, to answer a slightly different question: how to choose a small set of sensors with a relatively hight entropy (hence, independence)? How bad can one subset of sensors be compared to another of the same size? What is a good method to choose the better one?  The algorithm sequentially increases the size of the sensors set $I$ until its entropy estimate $\bw_n^{LZ}(I)$ does not grow. In a similar manner, one can choose empirical entropies. Due to the polymatroid properties we proved in the previous section, a bound on the performance compared to the optimum can be given.

The LZ parsing rule on an alphabet of size $\alpha$ can be implemented in $O(n\log \alpha)$ time (using an adequate tree and a binary enumeration of the alphabet). Hence, the complexity of {\sf GreedySelection} is $O(nK^3)$. \cite{nemhauser1978best} analyzed the performance of greedy schemes for submodular functions. As noted in \cite{shamaiah2010greedy} also for such algorithms, they achieve a factor of $1-\frac{1}{e}$ of the optimum.   

In practice, it might be beneficial to stop the algorithm if the entropy estimate does not grow above a certain threshold, to avoid steps which may include only a marginal improvement. In fact, this is exactly where the polymatroid properties we proved earlier kick in, and we have the following.
\begin{proposition}\label{prop. stop before end}
Assume Algorithm {\sf GreedySelection} is stopped after the first time $\bw_n^{LZ}(I)$ was incremented by less than some $\epsilon>0$. Then, for stationary and ergodic sources, the difference between the entropy of the currently selected subset of sensors and the entropy that could have been reached if the algorithm concluded is upper bounded by $K\epsilon+o(1)$. 
\end{proposition}  
\def\PROOFEARLYSTOP{
\begin{proof}
We wish to prove that if at some stage of the algorithm the improvement was some $\epsilon>0$, then no further step can improve by more than $\epsilon$, and hence the total improvement (till completion) is bounded by about $K\epsilon$.
To show this, we use the polymatroid axiom, and the fact that the LZ parsing rule estimates the entropy up to an additive estimation error of $o(1)$ (as $n$ increases).   

Let $\bw_n^{LZ}(I)$ be the estimated entropy at step $t$ of the algorithm, and $\bw_n^{LZ}(I \cup \{j^*\})$ be the estimated entropy at step $t+1$. We know that
\[
 \bw_n^{LZ}(I \cup \{j^*\})\leq \bw_n^{LZ}(I) + \epsilon.
\]
Also, for stationary and ergodic sources, with high probability, 
\[
|H(P^j, j \in \cS') - \bw_n^{LZ}(\cS')| = o(1)
\]
for any subset $\cS'\subseteq \cS$. Assume that at step $t+2$ the algorithm added a sensor $j' \ne j^*$ to the set $I\cup \{j^*\}$, such that 
\[
 \bw_n^{LZ}(I \cup \{j^*\} \cup \{j'\}) - \bw_n^{LZ}(I\cup \{j^*\}) > \epsilon.
\]
In this case, we have
\begin{eqnarray*}
\bw_n^{LZ}(I \cup \{j'\}) &\ge& H(P^j, j \in I \cup \{j'\})-o(1)
\\
&=&H(P^j, j \in I \cup \{j^*\} \cup \{j'\}) - H(P^{j^*}|P^j, j \in I \cup \{j'\})-o(1)
\\
&\ge&H(P^j, j \in I \cup \{j^*\} \cup \{j'\}) - H(P^{j^*}|P^j, j \in I)-o(1)
\\
&=&H(P^j, j \in I \cup \{j^*\} \cup \{j'\})
\\
&&\qquad - H(P^j, j \in I\cup\{j^*\})+H(P^j, j \in I)-o(1)
\\
&\ge&\bw_n^{LZ}(I \cup \{j^*\} \cup \{j'\})
 - \bw_n^{LZ}(I\cup\{j^*\})+\bw_n^{LZ}(I)-o(1)
\\
&>&\epsilon+\bw_n^{LZ}(I)-o(1).
\end{eqnarray*}
Hence, selecting $j'$ instead of $j^*$ at step $t+1$ would have been a better choice, which contradicts the greedy nature of the algorithm: we assumed $j^*$ was selected to maximize $\bw_n^{LZ}(I \cup \{j\})$ over all possible $j$. As a result, no further step can improve by more than $\epsilon$, and since there are at most $K$ steps left, the proposition follows.
\end{proof}
}
\PROOFEARLYSTOP
\section{A Sensor Fusion Algorithm Via Exponential Weighting}\label{sec. alg}
In this section, we present an online algorithm for sensor fusion. In \cite{Vovk90}, Vovk considered a general set of experts and introduced the \emph{exponential weighting}
algorithm. In this algorithm, each expert is assigned a weight
according to its past performance. By decreasing the weight of poorly
performing experts, hence preferring the ones proved to perform well thus
far, one is able to compete with the best expert, having neither any
\emph{a priori} knowledge on the input sequence nor which expert will
perform the
best. This result was further extended in \cite{Littlestone_Warmuth94},
where various aspects of a ``weighted majority" algorithm were discussed. In \cite{Cesa-Bianchi97,Hauss_Kivi_Warm98,Cesa-Bianchi_Lugosi99}, lower bound on the
redundancy of any universal algorithm were given, including very general loss functions. It is important to
note that the exponential weighting algorithm assumes nothing on the
set of experts, neither their distribution in the space of all
possible experts nor their structure. Consequently, all the results
are of the ``worst case'' type. Additional results regarding a randomized algorithm for expert selection can be found in \cite{Gyorfi_Lugosi_Morvai99} and \cite{Vovk98}.  

The exponential weighting algorithm was found useful also in the lossy
source coding works of Linder and Lugosi \cite{Linder_Lugosi01}, Weissman
and Merhav \cite{Weiss_Mer02}, Gyorgy \emph{et}.\ \emph{al}.\
\cite{Gyorgy_Linder_Lugosi04} and the derivation of sequential strategies
for loss functions with memory \cite{Mer_Orden_Serou_Weinb02}. A common
method in these works is the alternation of experts only once every block
of input symbols, necessary to bear the price of this change (e.g.,
transmitting the description of the chosen quantizer
\cite{Linder_Lugosi01}-\cite{Gyorgy_Linder_Lugosi04}). A major drawback of all the above
algorithms is the need to compute the performance of each expert at every
time instant. In \cite{Gyorgy_Linder_Lugosi04}, though, Gyorgy
\emph{et}.\ \emph{al}.\ exploit the
structure of the experts (as they are all quantizers) to introduce an
algorithm which efficiently computes the performance (or an approximation of it) of each
expert at each stage.

In this work, we offer to use a sequential strategy similar to the one used for loss functions with memory \cite{Mer_Orden_Serou_Weinb02} and scanning of multidimensional data \cite{Cohen_Merhav_Weissman_I07,Cohen_Weissman_Merhav_II08} in order to weight the sensors and identify the best fused sensor. However, given a set of sensors $\cS$, our goal is to construct a \emph{new sensor}, $\hat{S}$, whose output depends on the outputs of the given sensors, yet its performance is better than the best sensor in the set $\cS$.
We call $\hat{S}$ a \emph{synthesised (fused) sensor}. Clearly, when the true target appearance sequence $x_1^n$ is known in advance, suggesting such a sensor is trivial. However, we are interested in an \emph{online} algorithm, which receives the sensors' outputs at each time instant $t$, together with their \emph{performance in the past} (calculated by having access to $x_{t'}$ for $t' < t$ or estimating it), and computes a synthesised output. We expect the sequence of synthesised outputs given by the algorithm at times $t=1,\ldots,n$ to have a lower cumulative loss than the best sensor in $\cS$, for \emph{any possible sequence $x_1^n$ and any set of sensors $\cS$}.

Towards this goal, we will define a parametric set of synthesised sensors. Once such a set is constructed, say $\cS_\Theta$ for some set of parameters $\Theta$ (that is, $|\Theta|$ possible new sensors), we will use the online algorithm to compete with the \emph{best sensor in $\cS \cup \cS_\Theta$}. Clearly, a good choice for $\cS_\Theta$ is such that on the one hand $|\cS \cup \cS_\Theta|$ is not too large, yet on the other hand $\cS_\Theta$ includes ``enough" good synthesised sensors, so the best sensor in $\cS \cup \cS_\Theta$ will indeed perform well.
\begin{example}
A simple example to be kept in mind is a case where the set of sensors, $S_1, \ldots, S_K$, has the property such that all under-estimate the probability that a target exists (for example, since each sensor measures a different aspect of the target, which might not be visible each time the target appears). In this case, a sensor $\hat{S}$ whose output at time $t$ is $\max_{j}\{p^j_t,1\leq j \leq K\}$ will have a much smaller cumulative loss $L_{\hat{S}}(x_1^n)$ compared to any individual sensor, $L_{S_j}(x_1^n)$. 
As a result, when designing families of synthesised sensors for such a set of $K$ sensors, one can think of a set synthesised family $\cS_m$, which includes, for example, all sensors of the type $\max\{p^{j_1}_t, \ldots, p^{j_m}_t\}$ for some subset $\{j_1,\ldots,j_m\} \subset \{1,\ldots K\}$. If the miss-detection probabilities of the sensors are not all equal, clearly some synthesised set of $m$ sensors will perform better than the others.

This example can be easily extended to a case where sensors either under-estimate or over-estimate. Following a single sensor will give a non-negligible error, while a simple median filter (sensor-wise) on a sufficiently large set of sensors might give asymptotically zero error. 
\end{example}
\subsection{Exponential Weighting for a Parametric Family of Sensors.}
Recall that for any time instant $t\leq n$, $L_{S_j}(x_1^t)$ denotes the intermediate normalized cumulative loss of sensor $S_j$. Hence, $tL_{S_j}(x_1^t)$ is simply the unnormalized cumulative loss until (and including) time instant $t$. For simplicity, we denote this loss by $L_{j,t}$. Furthermore, note that for each $1 \leq j\leq K+|\Theta|$, $L_{j,0}=0$. At each time instant $t$, the exponential weighing algorithm assigns each sensor $S_j \in \cS \cup \cS_\Theta$ a probability $P_t(j|\{L_{j,t}\}_{j=1}^{K+|\Theta|})$. That is, it assumes the cumulative losses of all sensors up to time $t$ are known. Then, at each time instant $t$, after computing $P_t(j|\{L_{j,t}\}_{j=1}^{K+|\Theta|})$, the algorithm selects a sensor in $\cS \cup \cS_\Theta$ according to that distribution. The selected sensor is used to compute the \emph{algorithm output at time $t+1$}, namely, the algorithm uses the selected sensor as the synthesised sensor $\hat{S}$ at time $t+1$. Note that this indeed results in a synthesised sensor, as even if it turns out that the best sensor at some time instant is in $\cS$, it is not necessarily always the same sensor, hence the algorithm output will probably not equal any fixed sensor for all time instances $1\leq t \leq n$. The suggested algorithm is summarized in Algorithm {\sf OnlineFusion} below.
\begin{figure*}[ht]
\begin{center}
\framebox{\parbox{\textwidth}{
{\bf Algorithm  \sf OnlineFusion \\}
\% Input: $K+|\Theta|$ sensors, $\cS \cup \cS_\Theta$; Data $x_1^n$, arriving sequentially. \\
\% Output: At each time instance, a synthesised sensor $\hat{S}\in \cS \cup \cS_\Theta$, chosen at random, such that the excess cumulative loss compared to the best synthesised sensor is almost surely asymptotically (in $n$) negligible (see Proposition \ref{prop. exponential main} and the discussion which follows).   
\begin{itemize}
\item Initialization:
\\
$W=K+|\Theta|$; $\forall_{1\leq j \leq K+|\Theta|} \quad L_j=0$, $P(j|\{\})=\frac{1}{W}$; $\eta= \sqrt{\frac{8\log(K+|\Theta|)}{n d_{max}^2}}$.
\item For each $t=1,\ldots,n$:
\begin{itemize}
\item Choose $\hat{S}$ according to $P(j|\{\})$.
\item For each $j=1,\ldots,K+|\Theta|$:
\begin{itemize}
\item $L_j \gets L_j+d(p^j_t,x_t)$.
\end{itemize}
\item $W \gets \sum_{j=1}^{K+|\Theta|}e^{-\eta L_j}$.
\item For each $j=1,\ldots,K+|\Theta|$:
\begin{itemize}
\item $P(j|\{\}) \gets \frac{e^{-\eta L_j}}{W}$.
\end{itemize}
\end{itemize}
\end{itemize}
}}
\end{center}
\end{figure*}
The main advantage in this algorithm is that, under mild conditions, the normalized cumulative loss of the synthesised sensor $\hat{S}$ it produces is approaching that of the best sensor in $\cS \cup \cS_\Theta$, hence it converges to the best synthesised sensor in a family of sensors, without knowing in advance which sensor that might be. By the standard analysis of exponential weighing, the following proposition holds.
\begin{proposition}\label{prop. exponential main}
For any sequence $x_1^n$, any set of sensors $\cS$ of size $K$ and any set of synthesised sensors $\cS_\Theta$, the expected performance of Algorithm  {\sf OnlineFusion} is given by
$E[L_{\hat{S}}(x_1^n)] \leq \min_{S \in \cS \cup \cS_\Theta}L_S(x_1^n) + d_{max}\sqrt{\frac{\log(K+|\Theta|)}{2n}}$,
where the expectation is over the randomized decisions in the algorithm and $d_{max}$ is some upper bound on the instantaneous loss.
\end{proposition}
For completeness, a proof is given in \ref{app. exponential main}.
As a result, as long as $\log(K+|\Theta|) = o(n)$ the synthesised sensor $\hat{S}$ has a vanishing redundancy compared to the best sensor in $\cS \cup \cS_\Theta$. This gives us an enormous freedom in choosing the parametrized set of sensors $\cS_\Theta$, and even sets whose size grows polynomially with the size of the data are acceptable.

The performance of the exponential weighting algorithm can be summarized as follows. For any set of stationary sources with probability measure $Q$, as long as the number of synthesised sensors does not grow exponentially with the data, we have
\[
\liminf_{n \rightarrow \infty}
E_{Q}EL_{\hat{S}}(X_1^n)
\leq
\liminf_{n \rightarrow \infty}\min_{S \in \cS \cup \cS_\Theta}
E_{Q}L_{S}(X_1^n),
\]
where the inner expectation in the left hand side is due to the possible randomization in $\hat{S}$. When the algorithm bases its decisions on independent drawings, we have 
\[
\lim_{n \to \infty}L_{\hat{S}}(x_1^n) \leq \lim_{n \to \infty}\min_{S \in \cS \cup \cS_\Theta}L_S(x_1^n)\]
almost surely (in terms of the randomization in the algorithm).
If, furthermore, the sources are strongly mixing, almost sure convergence in terms of the sources distribution is guaranteed as well \cite{Cohen_Merhav_Weissman_I07}: 
\[
\liminf_{n \rightarrow \infty}
L_{\hat{S}}(X_1^n)
\leq
\liminf_{n \rightarrow \infty}\min_{S \in \cS \cup \cS_\Theta}
L_{S}(X_1^n), \text{Q-a.s.}
\]
A by product of the algorithm is the set of weights it maintains while running. These weights are, in fact, good estimates of the \emph{sensors' reputation}. Moreover, such weights can help us make intelligent decisions for synthesised control and fine-tuning of the sensor selection process, namely, we are able to clearly see which families of synthesised sensors perform better, and within a family, which set of parameters should be described in higher granularity compared to the others (since sensors with these values perform well).  
Finally, note that this is a \emph{finite horizon} algorithm, since the optimal $\eta$ depends on the size of the data, $n$. One can loose the dependence on the size of the data easily by working with exponentially growing blocks of data.
\section{Results on Real and Artificial Data}
To validate the proposed methods in practice, simulations were carried out on both real and synthetic data. We present here some of the results.  

\begin{figure}[ht]
\includegraphics[scale=0.8,angle=90]{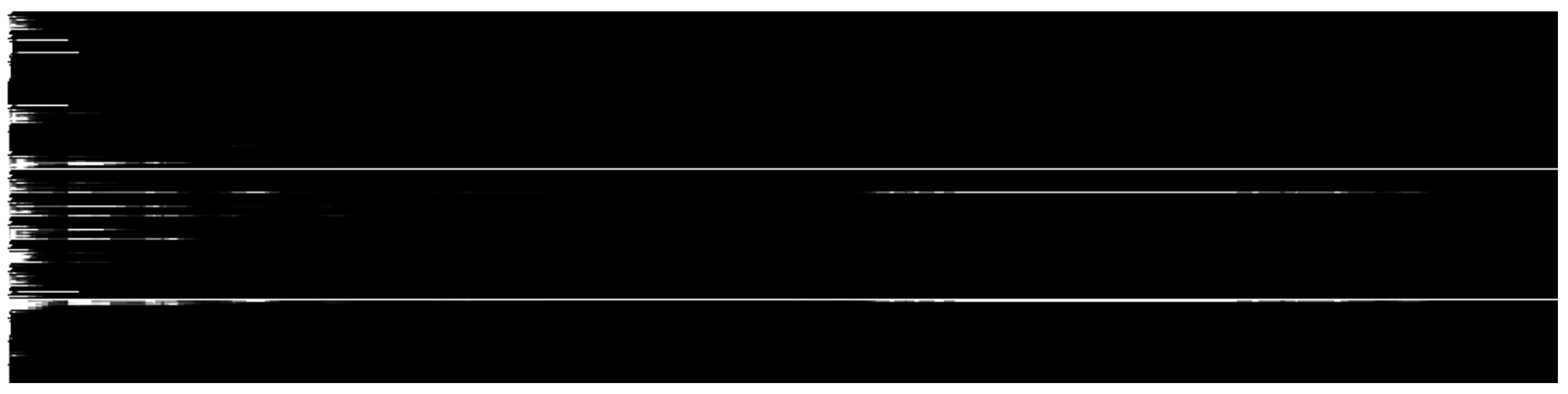}
 \label{fig. weights}
\caption{Weight vectors generated by {\sf OnlineFusion} for $240$ sensors - $15$ real sensors and $225$ synthesised ones. While at first (left column) all sensors have equal weight, as time evolves (towards the right) some sensors gain reputation (white color), while others loose it exponentially fast (dark color). Very soon, the two best sensors are identified.}
\end{figure}

\begin{figure}[ht]
\includegraphics[scale=0.6]{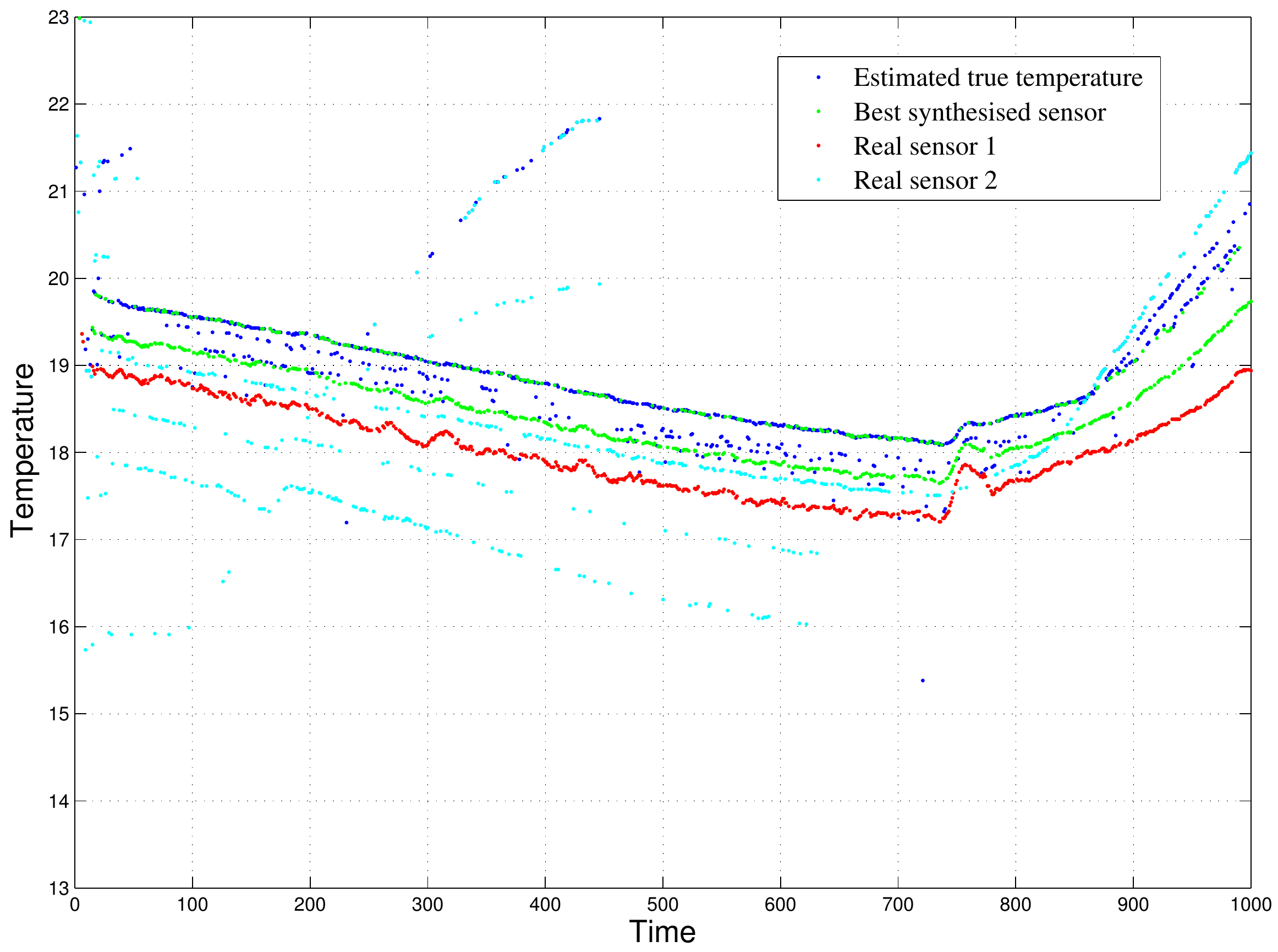}
\label{Temperature}
\caption{Real temperature data from the Intel Research Lab, Berkeley.}
\end{figure}

To demonstrate Algorithm {\sf OnlineFusion}, We used real sensors data collected from 54 sensors deployed in the Intel Berkeley Research Lab between February 28th and April 5th, 2004.\footnote{For details, see {\tt http://db.csail.mit.edu/labdata/labdata.html}.} To avoid too complex computations, we used only the first $15$ real sensors (corresponding to a wing in the lab) and artificially created from them $225$ fused (synthesised) sensors. For this basic example, the fused sensors were created by simply averaging the data of any two real sensors. Yet, the results clearly show how the best fused sensor outperforms the best real sensor, with very fast convergence times. Figure \ref{fig. weights} demonstrates the convergence of the weight vectors created by the algorithm. At start (left column), all weights are equal. Very fast, the two best sensors have a relatively high weight (approximately $0.5$), while the weight of the others decrease exponentially. Hence, the algorithm identifies the two best sensors very fast. The two best sensors are indeed synthesised ones, with the real sensors performing much worse. Note that there was no real data ($x_t$) for this sample. The real data was artificially created from \emph{all $15$ sensors} with a more complex function than simple average (first, artifacts where removed, then an average was taken). Thus, an average over simply two sensors, yet the best two sensors, outperforms any single one, and handles the artifacts in the data automatically.
Figure \ref{Temperature} depicts the data of two random real sensors (to avoid cluttering the graph), the artificially created true data $x_t$ and the best synthesised sensor.
\begin{table}[ht]
\caption{Entropy estimates for sensors from the Intel Research Lab, Berkeley.}
\label{table:realdata}
\begin{center}
\begin{small}
\begin{sc}
\begin{tabular}{lcc}
\hline
Method \quad & Sensor Numbers  \quad& Entropy Estimate   \\
\hline
Max.\ triplet & 1, 2, 8 & 4.0732 \\
Random & 15, 7, 2 & 2.9340 \\
Random &2,    4,   13&    3.3720 \\
Random &  10,   11,    6&    3.4966\\
Random &   2,   10,    8&    3.5630\\
Random &   5,    7,    1&    3.7798\\
Random &   7,    9,   15&    3.8290\\
Random &   1,    9,   14&    3.8511\\
Random &   2,    9,    4&    3.8528\\
Random & 11,   10,    1&    3.8570\\
Random & 12,    7,    2&    3.8730\\
Min.\ triplet & 15, 5, 7 & 2.4758 \\
\hline
\end{tabular}
\end{sc}
\end{small}
\end{center}
\end{table}

To demonstrate the greedy and random selection algorithms, we used the same data. Table \ref{table:realdata} includes the results. The entropy of the maximal triplet of sensors can be compared to that of random selections of triplets. Note that since many sensors are spread in a relatively small aria, there are several triplets which include an amount of information very close to the maximal (for a triplet). To get a sense of how correlated sensors can be, the entropy of a minimal triplet (also achieved by a greedy algorithm) is also depicted.

We also demonstrate the random sensor selection algorithm on artificial data. To do this, we artificially created randomized data for 5 independent sensors, and used them to create 5 additional depend ones, which are a function of the original  sensor. Sensors with even numbers are independent of each other, while sensors with odd number are linearly dependent on the even number sensors. Note that this is a very simplified model, which is included here only to demonstrate in practice the number of rounds the random selection algorithm requires in order to find an independent set. Furthermore, note that sensors depending on others and additional data may still be independent of each other, depending on the other sensors in the group. For example, if $P^1$ and $P^2$ are independent bits (with entropy $1$ each), and $P^3=P^1 \oplus P^2$, then $P^2$ and $P^3$ are still independent, with joint entropy $2$, while the three are dependent, with joint entropy $2$ as well. 

The algorithm then chose sets of $5$ sensors at random. Entropy estimates of the $5$ selected sensors are computed according to the joint first order probability estimate, that is, $H(T^n_{p^{j_1},\ldots,p^{j_5}})$, where $p^{j_1},\ldots,p^{j_5}$ is the data for the five selected sensors. It is easy to see from Table \ref{sample-table2} that $5$ independent sensors were drawn very fast, with $4$ out of $20$ trials succeeding. 
\begin{table}[ht]
\caption{Entropy estimate results of $20$ independent drawings of $5$ out of $10$ sensors.}
\label{sample-table2}
\vskip 0.15in
\begin{center}
\begin{small}
\begin{sc}
\begin{tabular}{cccc}
\hline
Draw Number \quad & Entropy Estimate  \quad& Draw Number  \quad& Entropy Estimate   \\
\hline
1 & 3.9938  & 11 & 3.9899  \\
2 & 3.9938 & 12 & 2.9966      \\
3 & 3.9938  &  13 & 4.9829 \\
4 & 3.9915  &  14 & 3.9938 \\
5 & 3.9938  &   15 & 3.9938  \\
6 & 2.9970  &  16 & 4.9829 \\
7 & 1.9976  &  17 & 4.9829  \\
8 & 4.9829  &  18 & 2.9966 \\
9 & 3.9895  &  19 & 2.9943   \\
10 & 3.9938  & 20 & 3.9938  \\
\hline
\end{tabular}
\end{sc}
\end{small}
\end{center}
\vskip -0.1in
\end{table}

\appendix
\section{Proof of Proposition \ref{prop. exponential main}}\label{app. exponential main}
We follow the analysis of exponential weighing, similar to \cite{Mer_Orden_Serou_Weinb02}. A similar analysis was also used in \cite{Cohen_Merhav_Weissman_I07}. In our setting, however, there is no notion of block size (so one can assume data is processed in blocks of size $1$).

For some $\eta > 0 $ define
\[
W_t = \sum_{j=1}^{K+|\Theta|}{e^{-\eta L_{j,t}}}
\]
and let the probability distribution assigned by the algorithm be
\begin{equation}
P_t(j|\{L_{j,t}\}_{j=1}^{K+|\Theta|}) = \frac{e^{-\eta L_{j,t}}}{W_t}, \quad 1\leq j\leq K+|\Theta|.
\end{equation}
We have
\begin{eqnarray}
\log\left(\frac{W_{n}}{W_0}\right) &=& \log\left(\sum_{j=1}^{K+|\Theta|}{e^{-\eta
L_{j,n}}}\right) - \log(K+|\Theta|)
\nonumber\\
&\geq& \log \left(\max_{1\leq j\leq K+|\Theta|}e^{-\eta L_{j,n}}\right)- \log(K+|\Theta|)
\nonumber\\
&=& -\eta \min_{1\leq j\leq K+|\Theta|}L_{j,n}- \log(K+|\Theta|)
\nonumber\\
&=& -\eta \min_{S \in \cS \cup \cS_\Theta}L_S(x_1^n)- \log(K+|\Theta|)
. \label{eq. lower bound on log of ratio}
\end{eqnarray}
Moreover,
\begin{eqnarray}
\log\left(\frac{W_{t+1}}{W_t}\right) &=& \log\left(
\frac{\sum_{j=1}^{K+|\Theta|}{e^{-\eta(L_{j,t}+d(p^j_{t+1},x_{t+1}))}}}{W_t}\right)
\nonumber\\
&=& \log \left(\sum_{j=1}^{K+|\Theta|}\frac{e^{-\eta L_{j,t}}}{W_t}e^{-\eta d(p^j_{t+1},x_{t+1})}\right)
\nonumber\\
&=& \log\left(
\sum_{j=1}^{K+|\Theta|}{P_t\left(j|\{L_{j,t}\}_{j=1}^{K+|\Theta|}\right)e^{-\eta
d(p^j_{t+1},x_{t+1})}}\right)
\nonumber\\
&\leq& -\eta
\sum_{j=1}^{K+|\Theta|}{P_t\left(j|\{L_{j,t}\}_{j=1}^{K+|\Theta|}\right)d(p^j_{t+1},x_{t+1})}+\frac{
d_{max}^2\eta^2}{8},
\nonumber\\
\end{eqnarray}
where the last inequality follows from assuming the distance function $d(\cdot,\cdot)$ is bounded by some $d_{max}$, hence $-\eta d(p^j_{t+1},x_{t+1})$ is in the range $[-\eta d_{max},0]$, and the extension to Hoeffding's inequality
given in \cite{Mer_Orden_Serou_Weinb02}, which asserts that for any random variable $Z$ taking values in a bounded interval of size $R$ and mean $\mu$ we have
\[
\log E[e^Z] \leq \mu + \frac{R^2}{8}.
\]
Thus,
\begin{eqnarray}
\log\left(\frac{W_{n}}{W_0}\right) &=&
\sum_{t=0}^{n-1}{\log\left(\frac{W_{t+1}}{W_t}\right)}
\nonumber\\
&\leq& -\eta \sum_{t=0}^{n-1}{\sum_{j=1}^{K+|\Theta|}
{P_t\left(j|\{L_{j,t}\}_{j=1}^{K+|\Theta|}\right)d(p^j_{t+1},x_{t+1})}}+\frac{
d_{max}^2\eta^2 n}{8}
\nonumber\\
&=& -\eta nE[L_{\hat{S}}(x_1^n)]+\frac{d_{max}^2\eta^2n}{8}, \label{eq. upper
bound on log of ratio}
\end{eqnarray}
where the expectation in \eqref{eq. upper bound on log of ratio} is with respect to the randomized choices the algorithm takes. Finally, from \eqref{eq. lower bound on log of ratio} and \eqref{eq. upper
bound on log of ratio}, we have, for any sequence $x_1^n$,
\begin{equation}\label{eq. bound on loss}
E[L_{\hat{S}}(x_1^n)] \leq \min_{S \in \cS \cup \cS_\Theta}L_S(x_1^n) + \frac{\log(K+|\Theta|)}{n \eta} + \frac{d_{max}^2 \eta}{8}.
\end{equation}
Since $\eta$ is any non-negative parameter, we may optimize the right hand side of \eqref{eq. bound on loss} with respect to $\eta$. The proposition follows by choosing
\[
\eta^* = \sqrt{\frac{8\log(K+|\Theta|)}{n d_{max}^2}}.
\]
%
\bibliographystyle{model1-num-names}
\bibliography{all_references}

\begin{thebibliography}{34}
\expandafter\ifx\csname natexlab\endcsname\relax\def\natexlab#1{#1}\fi
\providecommand{\bibinfo}[2]{#2}
\ifx\xfnm\relax \def\xfnm[#1]{\unskip,\space#1}\fi
\bibitem[{Slepian and Wolf(1973)}]{SlepianWolf73}
\bibinfo{author}{D.~Slepian}, \bibinfo{author}{J.~Wolf},
\newblock \bibinfo{title}{{N}oiseless coding of correlated information
  sources},
\newblock \bibinfo{journal}{IEEE Trans. Inform. Theory} \bibinfo{volume}{19}
  (\bibinfo{year}{1973}) \bibinfo{pages}{471--480}.
\bibitem[{Ziv and Lempel(1978)}]{Ziv_Lemp78}
\bibinfo{author}{J.~Ziv}, \bibinfo{author}{A.~Lempel},
\newblock \bibinfo{title}{{C}ompression of individual sequences via
  variable-rate coding},
\newblock \bibinfo{journal}{IEEE Trans. Inform. Theory} \bibinfo{volume}{IT-24}
  (\bibinfo{year}{1978}) \bibinfo{pages}{530--536}.
\bibitem[{Zozor et~al.(2005)Zozor, Ravier, and Buttelli}]{zozor2005lempel}
\bibinfo{author}{S.~Zozor}, \bibinfo{author}{P.~Ravier},
  \bibinfo{author}{O.~Buttelli},
\newblock \bibinfo{title}{On lempel-ziv complexity for multidimensional data
  analysis},
\newblock \bibinfo{journal}{Physica A: Statistical Mechanics and its
  Applications} \bibinfo{volume}{345} (\bibinfo{year}{2005})
  \bibinfo{pages}{285--302}.
\bibitem[{Blanc et~al.(2008)Blanc, Schmidt, Bonnier, Pezard, and
  Lesne}]{blanc2008quantifying}
\bibinfo{author}{J.~Blanc}, \bibinfo{author}{N.~Schmidt},
  \bibinfo{author}{L.~Bonnier}, \bibinfo{author}{L.~Pezard},
  \bibinfo{author}{A.~Lesne},
\newblock \bibinfo{title}{Quantifying neural correlations using lempel-ziv
  complexity},
\newblock in: \bibinfo{booktitle}{Neurocomp}.
\bibitem[{Hall and Llinas(1997)}]{hall1997introduction}
\bibinfo{author}{D.~Hall}, \bibinfo{author}{J.~Llinas},
\newblock \bibinfo{title}{An introduction to multisensor data fusion},
\newblock \bibinfo{journal}{Proceedings of the IEEE} \bibinfo{volume}{85}
  (\bibinfo{year}{1997}) \bibinfo{pages}{6--23}.
\bibitem[{Sasiadek(2002)}]{sasiadek2002sensor}
\bibinfo{author}{J.~Sasiadek},
\newblock \bibinfo{title}{Sensor fusion},
\newblock \bibinfo{journal}{Annual Reviews in Control} \bibinfo{volume}{26}
  (\bibinfo{year}{2002}) \bibinfo{pages}{203--228}.
\bibitem[{Waltz(1986)}]{waltzdata}
\bibinfo{author}{E.~Waltz},
\newblock \bibinfo{title}{Data fusion for c3i: A tutorial},
\newblock \bibinfo{journal}{Command, Control, Communications Intelligence (C3I)
  Handbook}  (\bibinfo{year}{1986}) \bibinfo{pages}{217--226}.
\bibitem[{Karger(1993)}]{karger1993random}
\bibinfo{author}{D.~Karger},
\newblock \bibinfo{title}{Random sampling in matroids, with applications to
  graph connectivity and minimum spanning trees},
\newblock in: \bibinfo{booktitle}{Foundations of Computer Science, 1993.
  Proceedings., 34th Annual Symposium on}, \bibinfo{organization}{IEEE}, pp.
  \bibinfo{pages}{84--93}.
\bibitem[{Berger et~al.(2004)Berger, Gritzmann, and
  de~Vries}]{berger2004minimum}
\bibinfo{author}{F.~Berger}, \bibinfo{author}{P.~Gritzmann},
  \bibinfo{author}{S.~de~Vries},
\newblock \bibinfo{title}{Minimum cycle bases for network graphs},
\newblock \bibinfo{journal}{Algorithmica} \bibinfo{volume}{40}
  (\bibinfo{year}{2004}) \bibinfo{pages}{51--62}.
\bibitem[{Vovk(1990)}]{Vovk90}
\bibinfo{author}{V.~G. Vovk},
\newblock \bibinfo{title}{{Aggregating strategies}},
\newblock \bibinfo{journal}{Proc. 3rd Annu. Workshop Computational Learning
  Theory, \normalfont{San Mateo, CA}}  (\bibinfo{year}{1990})
  \bibinfo{pages}{372--383}.
\bibitem[{Jeon and Landgrebe(1999)}]{jeon1999decision}
\bibinfo{author}{B.~Jeon}, \bibinfo{author}{D.~Landgrebe},
\newblock \bibinfo{title}{Decision fusion approach for multitemporal
  classification},
\newblock \bibinfo{journal}{Geoscience and Remote Sensing, IEEE Transactions
  on} \bibinfo{volume}{37} (\bibinfo{year}{1999}) \bibinfo{pages}{1227--1233}.
\bibitem[{Yu and Sycara(2006)}]{yu2006learning}
\bibinfo{author}{B.~Yu}, \bibinfo{author}{K.~Sycara},
\newblock \bibinfo{title}{Learning the quality of sensor data in distributed
  decision fusion},
\newblock in: \bibinfo{booktitle}{Information Fusion, 9th International
  Conference on}, \bibinfo{organization}{IEEE}, pp. \bibinfo{pages}{1--8}.
\bibitem[{Polikar et~al.(2006)Polikar, Parikh, and
  Mandayam}]{polikar2006multiple}
\bibinfo{author}{R.~Polikar}, \bibinfo{author}{D.~Parikh},
  \bibinfo{author}{S.~Mandayam},
\newblock \bibinfo{title}{Multiple classifier systems for multisensor data
  fusion},
\newblock in: \bibinfo{booktitle}{Sensors Applications Symposium, 2006.
  Proceedings of the 2006 IEEE}, pp. \bibinfo{pages}{180--184}.
\bibitem[{Oxley(1992)}]{Oxley92}
\bibinfo{author}{J.~G. Oxley}, \bibinfo{title}{Matroid Theory},
  \bibinfo{publisher}{Oxford Univ. Press}, \bibinfo{address}{Oxford, U.K.},
  \bibinfo{year}{1992}.
\bibitem[{Yeung(2002)}]{Yeung02}
\bibinfo{author}{R.~W. Yeung}, \bibinfo{title}{A First Course in Information
  Theory}, \bibinfo{publisher}{Springer}, \bibinfo{year}{2002}.
\bibitem[{Chan and Grant(2008)}]{Chan_Grant08}
\bibinfo{author}{T.~H. Chan}, \bibinfo{author}{A.~Grant},
\newblock \bibinfo{title}{{D}ualities between entropy functions and network
  codes},
\newblock \bibinfo{journal}{IEEE Trans. Inform. Theory} \bibinfo{volume}{54}
  (\bibinfo{year}{2008}) \bibinfo{pages}{4470--4487}.
\bibitem[{Mat{\'u}{\v{s}}(2007)}]{Matus07}
\bibinfo{author}{F.~Mat{\'u}{\v{s}}},
\newblock \bibinfo{title}{{T}wo constructions on limits of entropy functions},
\newblock \bibinfo{journal}{IEEE Trans. Inform. Theory} \bibinfo{volume}{53}
  (\bibinfo{year}{2007}) \bibinfo{pages}{320--330}.
\bibitem[{Cover and Thomas(2006)}]{cover2006elements}
\bibinfo{author}{T.~Cover}, \bibinfo{author}{J.~Thomas},
  \bibinfo{title}{Elements of information theory}, \bibinfo{publisher}{Wiley},
  \bibinfo{year}{2006}.
\bibitem[{Savari(1997)}]{savari1997redundancy}
\bibinfo{author}{S.~Savari},
\newblock \bibinfo{title}{Redundancy of the lempel-ziv incremental parsing
  rule},
\newblock \bibinfo{journal}{Information Theory, IEEE Transactions on}
  \bibinfo{volume}{43} (\bibinfo{year}{1997}) \bibinfo{pages}{9--21}.
\bibitem[{Lezaud(1998)}]{lezaud1998chernoff}
\bibinfo{author}{P.~Lezaud},
\newblock \bibinfo{title}{Chernoff-type bound for finite markov chains},
\newblock \bibinfo{journal}{The Annals of Applied Probability}
  \bibinfo{volume}{8} (\bibinfo{year}{1998}) \bibinfo{pages}{849--867}.
\bibitem[{Shamaiah et~al.(2010)Shamaiah, Banerjee, and
  Vikalo}]{shamaiah2010greedy}
\bibinfo{author}{M.~Shamaiah}, \bibinfo{author}{S.~Banerjee},
  \bibinfo{author}{H.~Vikalo},
\newblock \bibinfo{title}{Greedy sensor selection: Leveraging submodularity},
\newblock in: \bibinfo{booktitle}{Decision and Control (CDC), 49th IEEE
  Conference on}, pp. \bibinfo{pages}{2572--2577}.
\bibitem[{Nemhauser and Wolsey(1978)}]{nemhauser1978best}
\bibinfo{author}{G.~Nemhauser}, \bibinfo{author}{L.~Wolsey},
\newblock \bibinfo{title}{Best algorithms for approximating the maximum of a
  submodular set function},
\newblock \bibinfo{journal}{Mathematics of Operations Research}
  (\bibinfo{year}{1978}) \bibinfo{pages}{177--188}.
\bibitem[{Littlestone and Warmuth(1994)}]{Littlestone_Warmuth94}
\bibinfo{author}{N.~Littlestone}, \bibinfo{author}{M.~K. Warmuth},
\newblock \bibinfo{title}{{T}he weighted majority algorithm},
\newblock \bibinfo{journal}{Inform. Comput.} \bibinfo{volume}{108}
  (\bibinfo{year}{1994}) \bibinfo{pages}{212--261}.
\bibitem[{Cesa-Bianchi et~al.(1997)Cesa-Bianchi, Freund, Haussler, Helmbold,
  Schapire, and Warmuth}]{Cesa-Bianchi97}
\bibinfo{author}{N.~Cesa-Bianchi}, \bibinfo{author}{Y.~Freund},
  \bibinfo{author}{D.~Haussler}, \bibinfo{author}{D.~P. Helmbold},
  \bibinfo{author}{R.~E. Schapire}, \bibinfo{author}{M.~K. Warmuth},
\newblock \bibinfo{title}{{H}ow to use expert advice},
\newblock \bibinfo{journal}{Journal of the ACM} \bibinfo{volume}{44(3)}
  (\bibinfo{year}{1997}) \bibinfo{pages}{427--485}.
\bibitem[{Haussler et~al.(1998)Haussler, Kivinen, and
  Warmuth}]{Hauss_Kivi_Warm98}
\bibinfo{author}{D.~Haussler}, \bibinfo{author}{J.~Kivinen},
  \bibinfo{author}{M.~K. Warmuth},
\newblock \bibinfo{title}{{S}equential prediction of individual sequences under
  general loss functions},
\newblock \bibinfo{journal}{IEEE Trans. on Information Theory}
  \bibinfo{volume}{44} (\bibinfo{year}{1998}) \bibinfo{pages}{1906--1925}.
\bibitem[{Cesa-Bianchi and Lugosi(1999)}]{Cesa-Bianchi_Lugosi99}
\bibinfo{author}{N.~Cesa-Bianchi}, \bibinfo{author}{G.~Lugosi},
\newblock \bibinfo{title}{{O}n prediction of individual sequences},
\newblock \bibinfo{journal}{The Annals of Statistics} \bibinfo{volume}{27}
  (\bibinfo{year}{1999}) \bibinfo{pages}{1865--1895}.
\bibitem[{Gyorfi et~al.(1999)Gyorfi, Lugosi, and
  Morvai}]{Gyorfi_Lugosi_Morvai99}
\bibinfo{author}{L.~Gyorfi}, \bibinfo{author}{G.~Lugosi},
  \bibinfo{author}{G.~Morvai},
\newblock \bibinfo{title}{{A} simple randomized algorithm for sequential
  prediction of ergodic time series},
\newblock \bibinfo{journal}{IEEE Trans. Inform. Theory} \bibinfo{volume}{45}
  (\bibinfo{year}{1999}) \bibinfo{pages}{2642--2650}.
\bibitem[{Vovk(1998)}]{Vovk98}
\bibinfo{author}{V.~Vovk},
\newblock \bibinfo{title}{{A} game of prediction with expert advice},
\newblock \bibinfo{journal}{Journal of Computer and System Sciences}
  \bibinfo{volume}{56} (\bibinfo{year}{1998}) \bibinfo{pages}{153--173}.
\bibitem[{Linder and Lugosi(2001)}]{Linder_Lugosi01}
\bibinfo{author}{T.~Linder}, \bibinfo{author}{G.~Lugosi},
\newblock \bibinfo{title}{{A} zero-delay sequential scheme for lossy coding of
  individual sequences},
\newblock \bibinfo{journal}{IEEE Trans. Inform. Theory} \bibinfo{volume}{47}
  (\bibinfo{year}{2001}) \bibinfo{pages}{2533--2538}.
\bibitem[{Weissman and Merhav(2002)}]{Weiss_Mer02}
\bibinfo{author}{T.~Weissman}, \bibinfo{author}{N.~Merhav},
\newblock \bibinfo{title}{{O}n limited-delay lossy coding and filtering of
  individual sequences},
\newblock \bibinfo{journal}{IEEE Trans. Inform. Theory} \bibinfo{volume}{48}
  (\bibinfo{year}{2002}) \bibinfo{pages}{721--733}.
\bibitem[{Gyorgy et~al.(2004)Gyorgy, Linder, and
  Lugosi}]{Gyorgy_Linder_Lugosi04}
\bibinfo{author}{A.~Gyorgy}, \bibinfo{author}{T.~Linder},
  \bibinfo{author}{G.~Lugosi},
\newblock \bibinfo{title}{{E}fficient adaptive algorithms and minimax bounds
  for zero-delay lossy source coding},
\newblock \bibinfo{journal}{IEEE Trans. Signal Processing} \bibinfo{volume}{52}
  (\bibinfo{year}{2004}) \bibinfo{pages}{2337--2347}.
\bibitem[{Merhav et~al.(2002)Merhav, Ordentlich, Seroussi, and
  Weinberger}]{Mer_Orden_Serou_Weinb02}
\bibinfo{author}{N.~Merhav}, \bibinfo{author}{E.~Ordentlich},
  \bibinfo{author}{G.~Seroussi}, \bibinfo{author}{M.~J. Weinberger},
\newblock \bibinfo{title}{{O}n sequential strategies for loss functions with
  memory},
\newblock \bibinfo{journal}{IEEE Trans. Inform. Theory} \bibinfo{volume}{48}
  (\bibinfo{year}{2002}) \bibinfo{pages}{1947--1958}.
\bibitem[{Cohen et~al.(2007)Cohen, Merhav, and
  Weissman}]{Cohen_Merhav_Weissman_I07}
\bibinfo{author}{A.~Cohen}, \bibinfo{author}{N.~Merhav},
  \bibinfo{author}{T.~Weissman},
\newblock \bibinfo{title}{{S}canning and sequential decision making for
  multi-dimensional data - part {I}: the noiseless case},
\newblock \bibinfo{journal}{IEEE Trans. Inform. Theory} \bibinfo{volume}{53}
  (\bibinfo{year}{2007}) \bibinfo{pages}{3001--3020}.
\bibitem[{Cohen et~al.(2008)Cohen, Weissman, and
  Merhav}]{Cohen_Weissman_Merhav_II08}
\bibinfo{author}{A.~Cohen}, \bibinfo{author}{T.~Weissman},
  \bibinfo{author}{N.~Merhav},
\newblock \bibinfo{title}{{Scanning and Sequential Decision Making for
  Multidimensional Data - Part II: The Noisy Case}},
\newblock \bibinfo{journal}{IEEE Transactions on Information Theory}
  \bibinfo{volume}{54} (\bibinfo{year}{2008}) \bibinfo{pages}{5609--5631}.

\end{thebibliography}

\end{document}